\documentclass[final, journal, 9pt, ]{IEEEtran}

\usepackage{cite}
\usepackage{amsmath,amssymb,amsfonts}
\usepackage{graphicx}
\usepackage{textcomp}
\usepackage{xcolor}
\usepackage{caption}
\usepackage{algorithm}
\usepackage[noend]{algpseudocode}
\usepackage{scalerel}
\usepackage{subcaption}
\usepackage[nolist]{acronym}

\usepackage{tikz, circuitikz}
\usetikzlibrary{arrows,decorations.pathmorphing,backgrounds,positioning,fit,petri,quotes}
\usepackage{import}

\usetikzlibrary{svg.path}
\definecolor{orcidlogocol}{HTML}{A6CE39}
\tikzset{
  orcidlogo/.pic={
    \fill[orcidlogocol] svg{M256,128c0,70.7-57.3,128-128,128C57.3,256,0,198.7,0,128C0,57.3,57.3,0,128,0C198.7,0,256,57.3,256,128z};
    \fill[white] svg{M86.3,186.2H70.9V79.1h15.4v48.4V186.2z}
                 svg{M108.9,79.1h41.6c39.6,0,57,28.3,57,53.6c0,27.5-21.5,53.6-56.8,53.6h-41.8V79.1z M124.3,172.4h24.5c34.9,0,42.9-26.5,42.9-39.7c0-21.5-13.7-39.7-43.7-39.7h-23.7V172.4z}
                 svg{M88.7,56.8c0,5.5-4.5,10.1-10.1,10.1c-5.6,0-10.1-4.6-10.1-10.1c0-5.6,4.5-10.1,10.1-10.1C84.2,46.7,88.7,51.3,88.7,56.8z};
  }
}
\newcommand\orcidicon[1]{\href{https://orcid.org/#1}{\mbox{\scalerel*{
\begin{tikzpicture}[yscale=-1,transform shape]
\pic{orcidlogo};
\end{tikzpicture}
}{|}}}}

\usepackage{hyperref}
\hypersetup{
colorlinks=true, %
linktoc=all,     %
linkcolor=black,  %
}

\def\BibTeX{{\rm B\kern-.05em{\sc i\kern-.025em b}\kern-.08em
    T\kern-.1667em\lower.7ex\hbox{E}\kern-.125emX}}

\title{State Characterisation of Self-Directed Channel Memristive Devices}

\author{\IEEEauthorblockN{
D\'aniel Hajt\'o \orcidicon{0000-0002-5815-2256}\IEEEauthorrefmark{1},
Waleed El-Geresy \orcidicon{0000-0002-4016-6078}\IEEEauthorrefmark{2},
Deniz G\"und\"uz \orcidicon{0000-0002-7725-395X}\IEEEauthorrefmark{2},
Gy\"orgy Cserey \orcidicon{0000-0002-6836-1502}\IEEEauthorrefmark{1}}

\IEEEauthorblockA{P\'azm\'any P\'eter Catholic University\IEEEauthorrefmark{1}, Imperial College London\IEEEauthorrefmark{2}}

}

\begin{document}

\maketitle

\thispagestyle{plain}
\pagestyle{plain}

\begin{abstract}

    Knowing how to reliably use memristors as information storage devices is crucial not only to their role as emerging memories, but also for their application in neural network acceleration and as components of novel neuromorphic systems. In order to better understand the dynamics of information storage on memristors, it is essential to be able to characterise and measure their state. To this end, in this paper we propose a general, physics-inspired modelling approach for characterising the state of \ac{SDC} memristors. Additionally, to enable the identification of the proposed state from device data, we introduce a noise-aware approach to the minimum-variance estimation of the state from voltage and current pairs.

\end{abstract}

\acresetall

\section{Introduction}
\label{sec:introduction}

Memristors are a class of passive electronic device usually characterised by their ability to switch resistive state. First theoretically formulated by Leon Chua in 1971~\cite{chuaMemristorMissingCircuit1971}, they are devices whose resistive state can be modified by applying a voltage (voltage-controlled memristors) or a current (current or charge -controlled memristors).

Memristors were originally hypothetical devices, theorised as a class of passive electronic device with a non-differential relationship between the electromagnetic flux in the device and the electronic charge. This formulation was based on the devices being the missing piece needed to complete a ``periodic-table'' of passive electronic devices, consisting of resistors, capacitors, and inductors, each defined as connecting a pair of electronic state variables among \(v\), \(\frac{dv}{dt}\), \(i\), and \(\frac{di}{dt}\), through a non-differential relationship. Since their theoretical formulation, the term memristor has been used to describe a number of electronic devices~\cite{strukovMissingMemristorFound2008}, despite not necessarily strictly satisfying the original definition, and would be more correctly referred to as ``memristive systems''~\cite{chuaMemristiveDevicesSystems1976}.

This loose use of the term memristor is prevalent in the literature, with the term ``memristor'' often used to refer to any kind of resistive switching memory. Indeed, in 2011, Chua lent support to the idea of redefining memristors to include such devices~\cite{chuaResistanceSwitchingMemories2011}. This marked a transition from a mechanistic definition of memristors, to one based on their behaviour - namely, the characteristic pinched hysteresis loop indicative of a voltage or current-modulated instantaneous resistance - this instantaneous resistance often thought of as the state variable of the devices. For the purposes of simplicity, we will use the term ``memristor'' to refer to such resistive switching memories in this paper, since no real device satisfying the requirements of the original theoretical framework has yet been discovered.

One key feature of memristors - a consequence of the resistive switching property - is their ability to co-locate the processing and storage of information, circumventing the Von-Neumann bottleneck \cite{vonneumannFirstDraftReport1993}. This has the potential to enable for more energy-efficient storage and computation. They have hence emerged as promising candidates for a variety of applications including multilevel analogue data storage, where they offer potential benefits in terms of power consumption and storage density; energy-efficient deep neural network acceleration; and - having been shown early on to be theoretically capable of modelling the dynamics of biological neurons \cite{chuaMemristiveDevicesSystems1976} - as neuromorphic computing elements. For all these applications, it is important to have an understanding of the nature of information storage on the devices. The simplistic assumption that the memristor behaves as a resistor with memory is often unreasonable for complex devices that may exhibit non-linear characteristics. For example, non-linear VI characteristics can be evident \cite{wangAccountingMemristorIV2021}, and in general, the relationship between the device state and the instantaneous resistance is expressible through a (potentially non-linear) resistance readout equation \cite{el-geresyEventBasedSimulationStochastic2024}. A fundamental problem is therefore that of identifying stable state variables that can be used to describe the devices and characterise their state at each instant, without restrictions on the magnitude of voltage or current input variables, by making use of measurements.
Different states can also exhibit differing degrees of metastability, with some states being more stable than others, which can be modelled through the phenomenon of stochastic resistive drift \cite{el-geresyEventBasedSimulationStochastic2024, el-geresyDelayConditionedGenerative2024}. 

Various state models and associated state transition models have been proposed for different types of memristive devices, with each model having a different level of generality and physical grounding. Early on, \cite{strukovMissingMemristorFound2008} proposed a simple width-modulated switching model, with Ohmic conduction, to describe the behaviour of titanium oxide memristors, which assumed that the state variable was equivalent to the instantaneous resistance. Late models made use of exponential state VI characteristics. \cite{yangMemristiveSwitchingMechanism2008} described the conduction behaviour in the titanium oxide memristor as being modelled by two terms: a \(\sinh\) term based on trap-assisted electron-tunnelling through a thin electronic barrier and a non-linear Schottky diode term.

A further challenge for state modelling is presented by the fact that different types of memristors may have different physical characteristics, making distinct state characterisation approaches necessary for each.
A number of types of memristors exist, usually being split across several broad categories. These include electronic devices such as Mott insulators~\cite{janodResistiveSwitchingMott2015} and devices based on electron trapping~\cite{leeLayerbylayerAssembledChargetrap2007}; devices based on a redox reaction occurring in an active layer of the device, including \ac{SDC} memristors; \ac{PCM} devices~\cite{wongPhaseChangeMemory2010}; and other types of memristive device, such as organic memristors~\cite{liuOrganicBiomimickingMemristor2016}, and magnetic (or spin-transfer torque/spintronic) memristors~\cite{wangSpintronicMemristorSpinTorqueInduced2009}.
There is a trade-off between tailoring a model to make it effective for a particular physical device, and its performance when applied to a wider range of devices that may not have the same physical characteristics or switching mechanisms.

A further, less obvious, problem associated with state models is also present; although a range of state models for different memristive devices exist, the process of identifying the memristive state from straightforward voltage and current measurements may be difficult, and the uncertainty associated with state estimates may be unclear, with noiseless measurements often being assumed, which may not be the case.

In this paper, we study the state characterisation of \ac{SDC} redox memristive devices, with the aim of identifying a stable state variable using a general approach that is motivated both by the physics of real \ac{SDC} devices, but also general enough to be applicable to a variety of memristive technologies, more accurately characterising the VI relation and associated state than in previous works.
In an effort to make the model useful in practical scenarios, we also aim to improve the process of identifying the proposed state variable from measurements, by introducing a state estimation approach which can estimate our proposed state from noisy voltage and current pairs, quantifying the uncertainty associated with the estimates. We derive our state model based on the physics of thin film devices and apply our state characterisation protocol to data collected from \ac{SDC} memristive devices.

\textbf{Notation} We use zero-indexed square bracket notation to denote indexing of a vector, for example \(x[0]\) denotes the first element of the vector \(x\). Subscript notation \(X_i\) is used to distinguish multiple i.i.d. random variables of the same distribution as the random variable \(X\).

\section{Background and Motivation}
\label{sec:background_motivation}

In this section, we give background on the specific devices which will feature in our experiments - \ac{SDC} memristors - as well as introducing previous work which has attempted to model the state of \ac{SDC} memristors specifically.

\subsection{Self-Directed Channel Memristors}

Devices based on redox reactions are prevalent in the literature, and so further differentiation is needed on the basis of the nature of their switching mechanism. We can categorise them first, based on the type of ions that modulate switching in the devices. Redox memristors with a switching mechanism that relies on the movement of oxygen vacancies are called either \ac{VCM} or \ac{TCM} devices~\cite{waserRedoxBasedResistiveSwitching2009}, depending on whether bias or temperature, respectively, primarily governs the switching\footnote{Both mechanisms may be simultaneously present in such devices, however one is usually considered dominant.}. Such devices usually switch their state through the formation or destruction of conductive filaments within their active substrate, but can also sometimes operate based on interface-type switching. In contrast to \ac{VCM} and \ac{TCM} devices, those based on the movement of metal ions are called \ac{ECM} memristors. \ac{ECM} devices rely on the formation of metal-ion filaments or conducting paths in the device to change the resistance. Several memristive technologies rely on the \ac{VCM} switching mechanism, including \ac{CBRAM}~\cite{yuCompactModelingConductingBridge2011} and \ac{SDC} memristors.

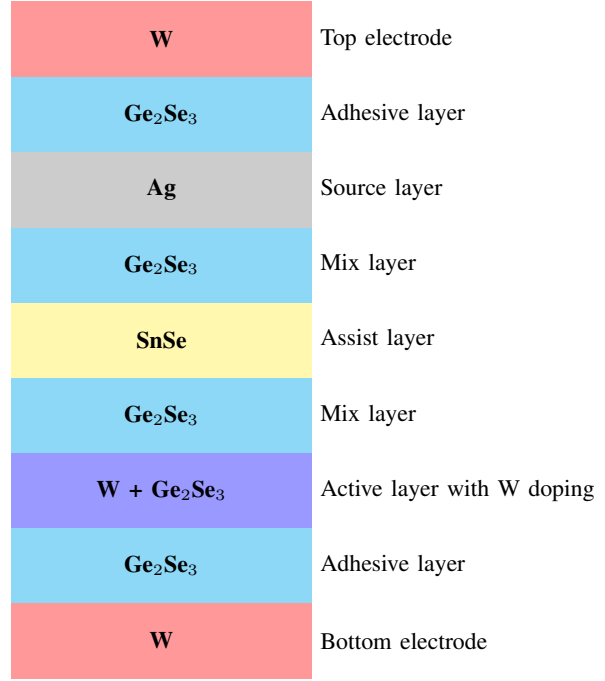
\begin{figure}
    \centering
    \begin{tikzpicture}

% Layer definitions
\newcommand{\layer}[5]{
  \node [rectangle, minimum width=4cm, minimum height=1cm, fill=#2, anchor=south] (#1) at (0, #3) {#4};
  \node [right of=#1, node distance=2cm, anchor=west] (#1-text) {#5};
}

% Draw the layers
\layer{bottom}{red!40}{0}{\textbf{W}}{Bottom electrode}
\layer{adhesive}{cyan!40}{1}{\textbf{Ge$_2$Se$_3$}}{Adhesive layer}
\layer{active}{blue!40}{2}{\textbf{W + Ge$_2$Se$_3$}}{Active layer with W doping}
\layer{mix1}{cyan!40}{3}{\textbf{Ge$_2$Se$_3$}}{Mix layer}
\layer{assist}{yellow!40}{4}{\textbf{SnSe}}{Assist layer}
\layer{mix2}{cyan!40}{5}{\textbf{Ge$_2$Se$_3$}}{Mix layer}
\layer{source}{gray!40}{6}{\textbf{Ag}}{Source layer}
\layer{adhesive2}{cyan!40}{7}{\textbf{Ge$_2$Se$_3$}}{Adhesive layer}
\layer{top}{red!40}{8}{\textbf{W}}{Top electrode}

\end{tikzpicture}
    \caption{The layer structure of the \ac{SDC} (Knowm) memristor used in our experiments.}
    \label{fig:knowm_memristor}
\end{figure}

\ac{SDC} memristors are thin film devices composed of a number of number of metal or metal compound layers~\cite{campbellSelfdirectedChannelMemristor2017}, doped to modify their switching properties, and have shown promise as being one of the few types of commercially available memristors. Knowm's W+\ac{SDC} memristor is one commercially available tungsten-doped \ac{SDC} device, with a structure as shown in Figure~\ref{fig:knowm_memristor}. The \(Ge_2Se_3\) layer, known as the ``active'' layer, is where changes related to switching typically manifest.
Devices are electroformed through the application of a positive potential to the top electrode, generating \(Sn\) ions from the \(SnSe\) layer and forcing them into the active layer, where a chemical reaction distorts the amorphous structure, forming conductive channels through the active layer that allow \(Ag+\) ions to carry charge through the layer. Switching is then primarily dictated by the movement of \(Ag+\) ions into or out of the conductive channels.
Modulation of the memrsitive state is achieved through creation and destruction of the \(Ag\) agglomeration sites within the insulating layer. The conduction occurs more easily when the agglomeration sites are more closely spaced and/or there is a higher concentration of Ag+ ions at these sites \cite{campbellSelfdirectedChannelMemristor2017}.

\subsection{State Modelling Approaches}

As mentioned in Section~\ref{sec:introduction}, models of memristive devices exist, with different advantages and use cases. There is a trade-off between two factors in state modelling, related to the generality of the modelling approach:

\begin{enumerate}
    \item Complexity and generality: how specific is the model to a particular device modelling setting?
    \item Physical grounding: does the model include inductive biases that are guided by knowledge of physical principles and specific device characteristics and operation?
\end{enumerate}

Although general models can be useful for describing general behaviour, they often omit useful inductive biases derived from specific physical modelling characteristics, that can tailor the models to specific devices and make them more accurate for those specific settings.

There are few models that describe the behaviour of \ac{SDC} memristors. One key model is the \ac{Generalised MSS}, a model developed by a team from Knowm, who commercialised the technology \cite{ nugentAHaHComputingFromMetastable2014, molterGeneralizedMetastableSwitch2016}.
The model incorporates both state and switching modelling components, imagining the devices to be a collection of metastable resistive switches whose individual states collectively control the variable overall state and resistance of the device.
Other works have adopted this modelling approach to improve the switching characteristics \cite{el-geresyEventBasedSimulationStochastic2024, el-geresyDelayConditionedGenerative2024} or to generalise the VI relation \cite{ostrovskiiStructuralParametricIdentification2021}, however, the accuracy of the model has yet to be experimentally verified.

In \cite{el-geresyEventBasedSimulationStochastic2024}, a state modelling approach was proposed that extended the approach given in \cite{nugentAHaHComputingFromMetastable2014}, proposing the separation of two distinct state variable types. The first is a state variable that has a many-to-one relationship (a readout equation) to the measured instantaneous device resistance - which we shall term the ``resistive state variable''. The second is a separate ``volatility state variable'' which affects the rates of transition of the resistive state variable, and can incorporate temporary effects and states that may change the switching dynamics, such as temperature and internal device structure.

\section{State Modelling}
\label{sec:state_modelling}

Any approach to state modelling first involves specifying what the ``state'' of the device actually refers to. As part of our modelling, we will concern ourselves only with the resistive state variable, and therefore, to avoid ambiguity, we will henceforth use ``state'' and ``state variable'' to refer to the resistive state alone.

The memristor modelling framework proposed in \cite{el-geresyEventBasedSimulationStochastic2024} assumes that the resistive state variable - which we denote \(x\) - is related, via a readout equation \(f\), to an instantaneous resistance \(r\). This implicitly assumes that that the shape of the VI characteristic for the memristor is linear.

\begin{align}
    f &: x \rightarrow r \nonumber \\
    i &= \frac{v}{r} \nonumber
\end{align}

However, the instantaneous VI characteristics of memristive devices associated with single resistive states are often non-linear \cite{wangAccountingMemristorIV2021}. This can be attributed to various material effects, such as the formation of Schottky diode barriers at film interfaces \cite{molterGeneralizedMetastableSwitch2016}.
Initial experiments we had conducted on \(Ge_2Se_3\) devices confirmed the characteristic of these devices was not linear, and appeared to be exponential in nature~\cite{hajtoRobustMemristorNetworks2019}. The VI characteristic was also asymmetric for positive and negative biases. %

For the purposes of more accurately modelling the state, we therefore propose a generalisation of the memristor modelling framework introduced in \cite{el-geresyEventBasedSimulationStochastic2024}, by instead assuming that the resistive state parameterises a \textit{function}, \(g_x\), that describes the relation between the current and the voltage.

\begin{align}
    i = g_{x}(v)
\end{align}

Note that the original readout equation formulation is a special case of the above, where:

\begin{align}
    g_{x}(i) = f(x) \cdot i
\end{align}

\begin{figure}
    \centering
    \includegraphics[width=\linewidth]{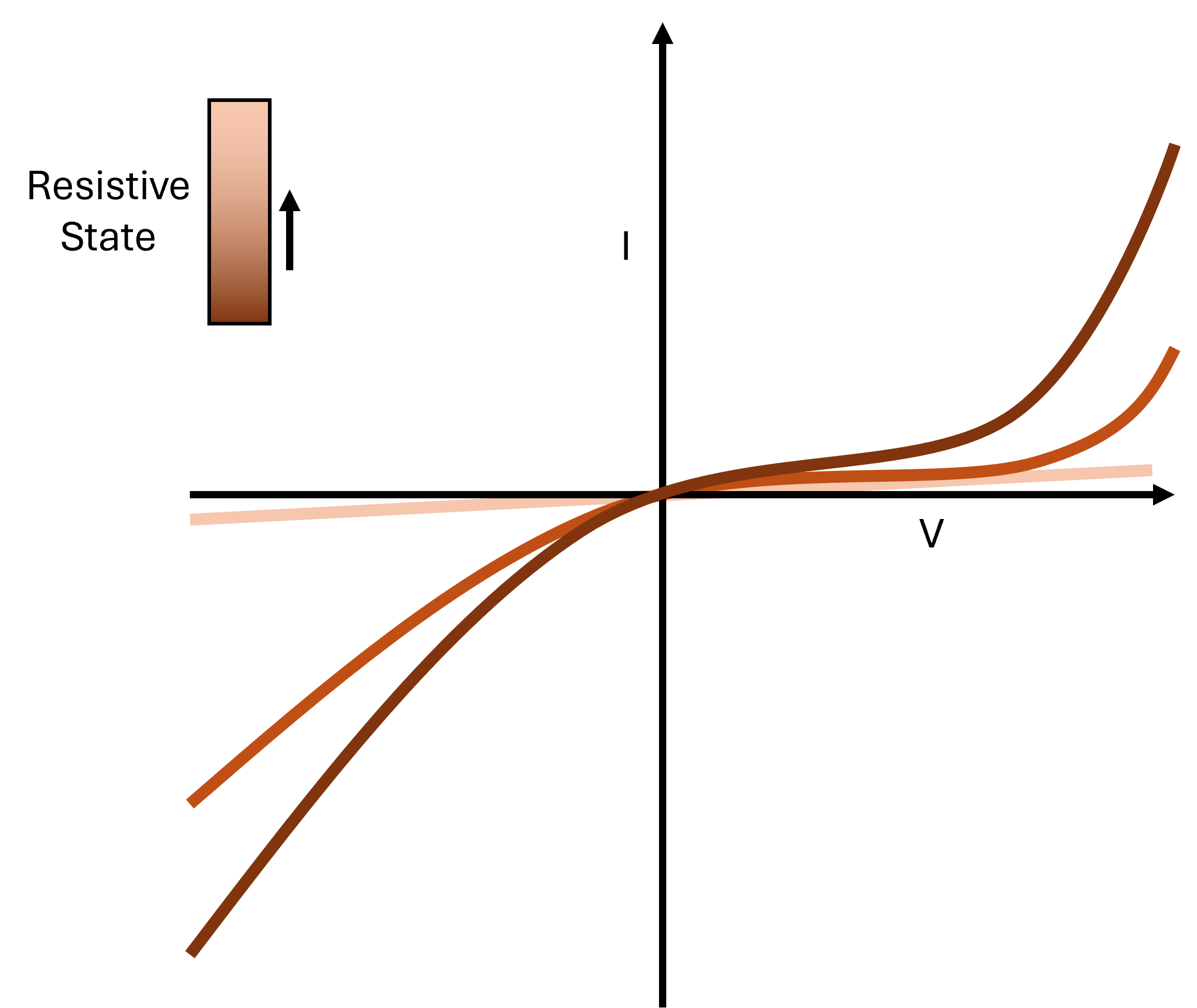}
    \caption{Each state of the memristor parameterises a VI characteristic, where each voltage input corresponds to a unique current output. A change of state is associated with a shift to a different set of voltage-current pairs.}
    \label{fig:family_of_curves}
\end{figure}

We make the assumption that the function \(g_x\) is one-to-one, that is, for a given resistive state, each voltage input will yield a single and unique current output. Thus, \(g_x\) presents a family of VI curves as shown in Figure~\ref{fig:family_of_curves}. The one-to-one nature also means the state can theoretically be determined from any given pair of voltage and current values, assuming noiseless measurement.

In this section, we detail our efforts to modify and extend the \ac{Generalised MSS} resistive state model, such  that it can be used to accurately characterise the resistive state of \ac{SDC} devices and model the associated family of non-linear VI characteristics, based on an understanding of the underlying physical mechanisms governing conduction in the devices.

\subsection{Physical Model}

In the case of \ac{SDC} memristive devices, we motivate our modelling approach based on a picture of the physical mechanisms underlying conduction.

In the case of \ac{SDC} memristors, the metal-semiconductor junction formed by \(Ag\) (silver) and \(Ge_2Se_3\) can be seen as giving rise to a Schottky diode-like IV characteristic. This behaviour is modelled using a variation of the Shockley diode equation. One can choose to include both a forward and reverse Schottky diode component, accounting for both the exponential and asymmetric nature of the IV characteristic.

\begin{figure}
    \centering
    %\documentclass{article}
%\usepackage{tikz}

%\begin{document}

\begin{tikzpicture}

    % Left Figure (Fewer Circles)
    \draw[thick, fill=blue!30] (0,0) rectangle (4,5); % Bounding box with blue fill

    % Define circle properties
    \def\rA{0.15} % Radius of first circle
    \def\rB{0.2} % Radius of second circle
    \def\xA{2}   % X coordinate of first circle center
    \def\yA{4}   % Y coordinate of first circle center
    \def\xB{0.8}   % X coordinate of second circle center
    \def\yB{2.5}   % Y coordinate of second circle center

    % Draw circles with black!70 color
    \filldraw[black!70] (\xA,\yA) circle (\rA);
    \filldraw[black!70] (\xB,\yB) circle (\rB);

    % Compute unit direction vector
    \pgfmathsetmacro{\dx}{\xB - \xA}
    \pgfmathsetmacro{\dy}{\yB - \yA}
    \pgfmathsetmacro{\dist}{sqrt(\dx*\dx + \dy*\dy)}
    \pgfmathsetmacro{\ux}{\dx / \dist}
    \pgfmathsetmacro{\uy}{\dy / \dist}

    % Compute start and end points of the shortened line
    \pgfmathsetmacro{\xAnew}{\xA + \ux * \rA}
    \pgfmathsetmacro{\yAnew}{\yA + \uy * \rA}
    \pgfmathsetmacro{\xBnew}{\xB - \ux * \rB}
    \pgfmathsetmacro{\yBnew}{\yB - \uy * \rB}

    % Distance Arrow (Left)
    \draw[thick, <->] (\xAnew, \yAnew) -- (\xBnew, \yBnew);

    % Other blobs with black!70 color
    \filldraw[black!70] (3,4.5) circle (0.10);
    \filldraw[black!70] (2,1) circle (0.25);

    % Top Cyan Box (Left Figure)
    \fill[cyan!40] (0,5) rectangle (4,6);

    % Bottom Cyan Box (Left Figure)
    \fill[cyan!40] (0,-1) rectangle (4,0);

    % Add "+" sign above the left figure
    \node at (2, 5.5) {\huge +};

    % Add "-" sign below the left figure
    \node at (2, -0.5) {\huge -};

    % Right Figure (More & Larger Circles)
    \draw[thick, fill=blue!30] (4.5,0) rectangle (8.5,5); % Bounding box with blue fill (shifted left by 1 unit)

    % Define circle properties
    \def\rC{0.25} % Radius of first circle (C)
    \def\rD{0.15} % Radius of second circle (D)
    \def\xC{6.5}   % X coordinate of first circle center
    \def\yC{4}   % Y coordinate of first circle center
    \def\xD{6}   % X coordinate of second circle center
    \def\yD{3}   % Y coordinate of second circle center

    % Draw circles with black!70 color
    \filldraw[black!70] (\xC,\yC) circle (\rC);
    \filldraw[black!70] (\xD,\yD) circle (\rD);

    % Compute unit direction vector
    \pgfmathsetmacro{\dx}{\xD - \xC}
    \pgfmathsetmacro{\dy}{\yD - \yC}
    \pgfmathsetmacro{\dist}{sqrt(\dx*\dx + \dy*\dy)}
    \pgfmathsetmacro{\ux}{\dx / \dist}
    \pgfmathsetmacro{\uy}{\dy / \dist}

    % Compute start and end points of the shortened line
    \pgfmathsetmacro{\xCnew}{\xC + \ux * \rC}
    \pgfmathsetmacro{\yCnew}{\yC + \uy * \rC}
    \pgfmathsetmacro{\xDnew}{\xD - \ux * \rD}
    \pgfmathsetmacro{\yDnew}{\yD - \uy * \rD}

    % Distance Arrow (Right)
    \draw[thick, <->] (\xCnew, \yCnew) -- (\xDnew, \yDnew);
    
    % Other blobs with black!70 color
    \filldraw[black!70] (5.3,2.5) circle (0.30);
    \filldraw[black!70] (7.5,4.5) circle (0.20);
    \filldraw[black!70] (6.5,1) circle (0.35);
    \filldraw[black!70] (7.7,2) circle (0.25);

    % Top Cyan Box (Right Figure)
    \fill[cyan!40] (4.5,5) rectangle (8.5,6);

    % Bottom Cyan Box (Right Figure)
    \fill[cyan!40] (4.5,-1) rectangle (8.5,0);

    % Add "+" sign above the right figure
    \node at (6.5, 5.5) {\huge +};

    % Add "-" sign below the right figure
    \node at (6.5, -0.5) {\huge -};

\end{tikzpicture}

%\end{document}
    \caption{Proposed physical model explaining the \ac{SDC} memristor's switching mechanism, with the gaps between silver ion agglomeration sites resulting in an electron tunnelling barrier conduction characteristic \cite{yangMemristiveSwitchingMechanism2008}. Increased density of hopping sites accelerates conduction.} %
    \label{fig:agglomeration}
\end{figure}
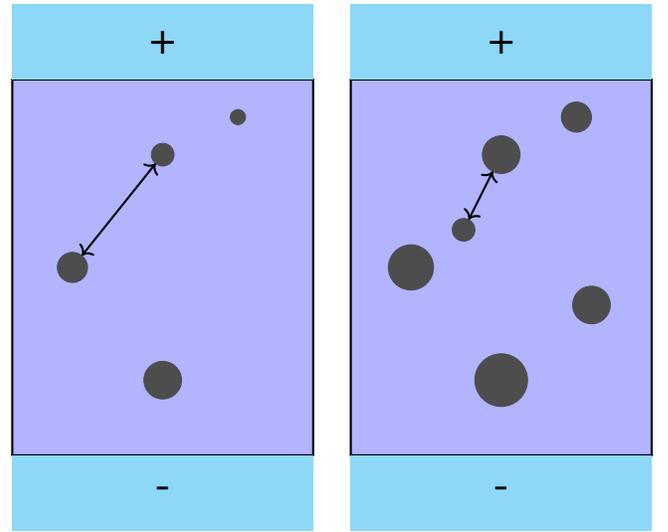

Simultaneously, the gaps between Ag+ agglomeration sites create a parallel conduction path. The collective effect of these gaps produces an electron tunnelling barrier behaviour, which is commonly modelled using a \(\sinh\) switching component. Figure~\ref{fig:agglomeration} depicts the model for the hypothetical conduction mechanism. Here, the modification of the internal structure through the change in concentration and size of \(Ag+\) ion sites is thought to affect the average gap between the conduction sites, consequently impacting the effective resistance of the active layer to the flow of charge, modulating the memristive state.

In summary, the conduction behaviour in the devices, and thus the state, can therefore be explained according to up to three factors:

\begin{itemize}
    \item Ohmic (resistive) behaviour of the active layer.
    \item A \(\sinh\) characteristic attributed to electron tunnelling at the metal-semiconductor junction \cite{strukovMissingMemristorFound2008, yangMemristiveSwitchingMechanism2008, molterGeneralizedMetastableSwitch2016}.
    \item Schottky diode-like characteristics at the metal-semiconductor junction.
\end{itemize}

\subsection{Ohmic Approximation and Generalised MSS Model}

Though the previously proposed Shockley diode physical model of the memristor at the metal-semiconductor junction may accurately represent conduction behaviour, it is also possible to assume the behaviour to instead be Ohmic as a first-order (linear) approximation. This reduces the modelling complexity, while retaining an approximation of the conduction characteristic.

Combining a forward and reverse exponential conduction component with a linear component yields a form similar to that of the \ac{Generalised MSS} model. An equivalent form of the conduction characteristic for the \ac{Generalised MSS} is given as follows:

\begin{align}
    i(x,v(t)) = x\cdot G_m \cdot v(t) + I_d(v(t)) \label{eq:gmss_characteristic}
\end{align}
where the diode conduction component is given as:
\begin{align}
    I_d(v(t)) = \alpha_1 \exp(\beta_1 v(t)) - \alpha_2\exp(-\beta_2 v(t)) \label{eq:gmss_diode}.
\end{align}

Note that we have above removed reference to \(N\) (the number of metastable switches), \(n\) the switches in the ON state, \(G_{\text{ON}}\), and \(G_{\text{OFF}}\) from the original model formulation, replacing them instead with a single parameterised conductance \(x \cdot G_m\) which models the \ac{Generalised MSS} memristive state. This is because we will not be modelling the switching behaviour itself and only the conduction characteristic.

\subsection{Proposed Model}

In similar spirit to \cite{molterGeneralizedMetastableSwitch2016}, we propose a modelling approach involving a parallel combination of a diode component, (\(I_d\)), given in Equation~\ref{eq:schockley_modified}, and an Ohmic component parameterised by \(G_m\).
In contrast to the \ac{Generalised MSS}, however, we apply two modifications to make the the model more physically plausible and to reduce the fitting error:

\begin{enumerate}
    \item We modify the diode component such that it is state dependent.
    \item We correct the diode equations such that they correspond to the characteristics of true Schottky diodes, to avoid non-zero crossing behaviour and to increase the model expressiveness.
\end{enumerate}

The first of these modifications is motivated by the idea that the Schottky diode characteristics are not independent of the memristive state. An increase in the conductivity of the substrate due to, for example, an increased concentration of hopping sites, will simultaneously decrease the effective barrier width of the Schottky diodes at the interface. We model this effect linearly, simply multiplying the state variable \(x\) by both the Ohmic conductance and the diode conduction components. We later show in Section~\ref{sec:results} that this method proves effective in minimising the fitting error:

\begin{align}
     i(x,v(t)) = x &\cdot (G_m \cdot v(t) + I_d(v(t)))
    \label{eq:modelling_equation}
\end{align}
where $G_m$ is a state-independent variable and $x$ is a parameter representing the memristor state.

Our second modification involves restoring the form of the diode equations given in \(I_d(v(t))\), such that non-zero-crossing models cannot be feasible. As given the \ac{Generalised MSS} proposes two independent exponential components with parameters \(\alpha_1\) and \(\alpha_2\). However, in order to ensure that the model maintains the zero-crossing property (a zero current is matched by a zero voltage in the conduction characteristic), we require \(\alpha_1=\alpha_2\). This is redundant, needlessly increasing the modelling complexity, and also reduces the model expressiveness by requiring the forward and reverse diode parameters to be equal. We therefore modify \(I_d(v(t))\) as follows:

\begin{align}
    I_d(v(t)) = \alpha_1 \cdot (e^{\beta_1 \cdot v(t)} - 1) + \alpha_2 \cdot (1 - e^{-\beta_2 \cdot v(t)})
    \label{eq:schockley_modified}
\end{align}

\section{Experimental Method}
\label{sec:experimental_method}

To verify the accuracy of our state model, we collected data for a range of states, assuming that each state corresponded to a one-to-one VI relation, as described in Section~\ref{sec:state_modelling}.

We conducted a series of experiments to measure the voltage and current characteristics of the memristors following the application of SET pulses of different magnitudes.

In our experiments, we used several waveform shapes to either program the device (change the device state) or to profile the device (measure the device state). The key operations are detailed here, which make use of two underlying waveform shapes, a square pulse and a triangular wave:

\begin{itemize}
    \item \textbf{RESET pulse}: We apply a square pulse of a fixed duration and a large negative magnitude to the device, parameterised by \(T_{reset}\) and \(A_{reset}\), respectively denoting the pulse duration and amplitude.
    \item \textbf{SET pulse}: We apply a square pulse of a fixed duration and a large positive magnitude to the device, parameterised by \(T_{set}\) and \(A_{set}\), respectively denoting the pulse duration and amplitude.
    \item \textbf{Read waveform}: We apply a small-magnitude triangular wave of a fixed magnitude to the device, parameterised by \(T_{read}\) and \(A_{read}\), respectively denoting the triangular wave period and amplitude.
\end{itemize}

For SET and RESET voltage waveforms that change the memristor state, we can consider there to be two dimensions: the total duration of the programming pulse, and the magnitude (which may possibly be a continuous function of time, resulting in different pulse shapes) of the voltage input. In order to simplify our analysis, we fixed the dimension of the duration of the SET and RESET pulses, and made the assumption that the magnitude should be constant over the duration of the pulse. A square pulse shape with a steep, but not negligible rising and falling edge was used comprising the first and last \(9\%\) of the period for the SET pulse, and the first and last \(16\%\) of the period for the RESET pulse. The non-zero duration of the rising and falling edges were used to gain more detailed information about the memristive state during the ramp up/down of the pulse magnitude, and also to minimise undesirable effects associated with sudden voltage magnitude changes.

In the case of the READ waveform, we use a small amplitude input value that was a magnitude sufficiently low as to be thought to have a negligible impact on the device state to measure the device resistance, and assumed that the memristor was in the same state when the associated current for this voltage was the same. A triangular waveform is used to minimise the time spent close to the maximum amplitude \(T_{read}\) and hence to minimise state changes, additionally yielding samples that are uniformly spaced over the given range of voltage inputs.

To collect our dataset, we first RESET the devices using a large negative magnitude pulse. We then read the devices using a triangular waveform to verify that the a high magnitude initialisation state has been achieved. We next SET the devices to bring them into a lower resistance state. Following this, we read the devices again to collect VI measurements for the given state. SET pulses of varying magnitudes were used to access different device states.

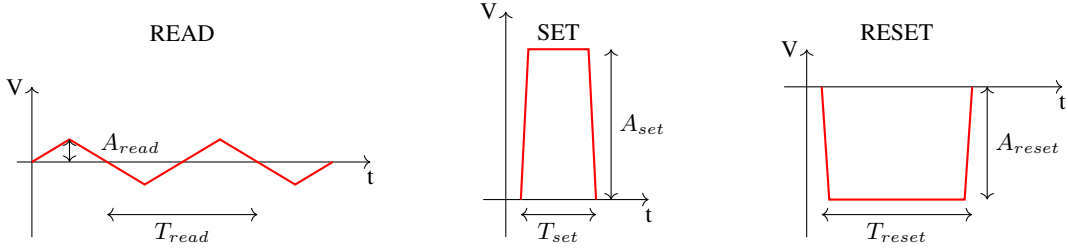
\begin{figure*}
    \centering
    \begin{tikzpicture}[scale = 1] 
    % Read phase
    \draw[->] (-0.2,0.5) -- (4.5,0.5) node[below] {t};
    \draw[->] (0,-0.5) -- (0,1.5) node[left] {V};
    \draw[thick, red] (0,0.5) -- (0.5,0.8) -- (1.5,0.2) -- (2.5,0.8) -- (3.5,0.2) -- (4,0.5);
    \draw[<->] (0.5,0.5) -- (0.5,0.8);
    \node[right] at (0.8,0.75) {$A_{read}$};
    \draw[<->] (1,-0.2) -- (3,-0.2);
    \node[below] at (2,-0.2) {$T_{read}$};
    \node[above] at (2,2) {READ};
    
    % Set phase
    \draw[->] (6,0) -- (8.2,0) node[below] {t};
    \draw[->] (6.3,-0.5) -- (6.3,2.5) node[left] {V};
    \draw[thick, red] (6.5,0) -- (6.6,2) -- (7.4,2) -- (7.5,0);
    \draw[<->] (7.7,0) -- (7.7,2);
    \node[right] at (7.7,1) {$A_{set}$};
    \draw[<->] (6.5,-0.2) -- (7.5,-0.2);
    \node[below] at (7,-0.2) {$T_{set}$};
    \node[above] at (7,2) {SET};
    
    % Reset phase
    \draw[->] (10,1.5) -- (13.7,1.5) node[below] {t};
    \draw[->] (10.3,-0.5) -- (10.3,2) node[left] {V};
    \draw[thick, red] (10.5,1.5) -- (10.6,0) -- (12.4,0) -- (12.5,1.5);
    \draw[<->] (12.7,0) -- (12.7,1.5);
    \node[right] at (12.7,0.75) {$A_{reset}$};
    \draw[<->] (10.5,-0.2) -- (12.5,-0.2);
    \node[below] at (11.5,-0.2) {$T_{reset}$};
    \node[above] at (11.5,2) {RESET};
\end{tikzpicture}

% read smaller
% reset larger
    \caption{Illustrative figures showing the READ, SET, and RESET waveforms used to measure and set/reset the memristors as part of our experiments. \(T_{read}\), \(T_{set}\), and \(T_{reset}\) are variables denoting the period of the waveforms, respectively. While, \(A_{read}\), \(A_{set}\), and \(A_{reset}\) denote the given amplitudes.}
    \label{fig:waveform}
\end{figure*}

\subsection{Measurement Equipment and Setup}

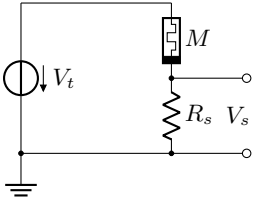
\begin{figure}[ht]
\centering
\begin{circuitikz}[scale = 1.0] 
\draw (0, 2) to[/tikz/circuitikz/bipoles/length=.75cm,V,v=\(V_{t}\),-*] (0, 0);
\draw (0,2) to[-] (2,2);
\draw (2, 2) to[/tikz/circuitikz/bipoles/length=.75cm,memristor,l=\(M\)] (2, 1);
\draw (2, 1) to[/tikz/circuitikz/bipoles/length=.75cm,R,l=\(R_{s}\),*-*] (2, 0);
\draw (2,1) to[short,-o] (3,1);
\draw (2.5,0.5) node[label={0:\(V_{s}\)}]{};
\draw (3,0) to[short,o-] (0,0);
\draw (0,0) node[ground]{} (0,-0.2);
\end{circuitikz}
\caption{The measurement circuit for the memristor. \(V_{\text{tot}}\) is the voltage measured across the entire circuit, while \(V_{\text{series}}\) is the voltage across the series resistor. From these measurements, we can determine the voltage across and current through the memristor.}
\label{fig:measurement_circuit}
\end{figure}

The voltage generation and data acquisition were done using a \textbf{Keysight InfiniiVision MSO-X 3054A} oscilloscope instrument. The measurement setup consists of a single ``Knowm SDC W'' memristor, in series with a \(100k\Omega\) resistor, \(R_{s}\), as shown in Figure~\ref{fig:measurement_circuit}. The main disadvantage of this measurement setup is that it does not allow for a constant voltage applied across the device. As the state of the memristor changes, the voltage-current relationship also changes, making it challenging to maintain a constant voltage across the device during measurements.

Voltage data are acquired at a ``high resolution" setting, which uses a temporal averaging to reach higher precision. The measurement results are 160 kSample/s data, but the actual sampling rate used internally by the oscilloscope was 2 GSample/s. Therefore, a datapoint is the average of 12,500 real samples.

\subsection{Experimental Data}

We used SET and RESET pulses to bring the devices into a representative set over the range of possible states. The devices were manually programmed using set pulses to bring them into states of varying magnitudes. Once the device was determined to be in a particular stable state according to an initial READ waveform cycle that was manually reviewed, a longer READ waveform was used to collect a set of VI measurements for the device in the given state.

We collected VI measurements for a variety of memristive device states, with a total of X separate state measurements. For each state measurement, we collected 4  READ waveform measurements, with each measurement consisting of 100 READ waveform cycles. The read waveform amplitude applied to the series circuit, shown in Figure~\ref{fig:measurement_circuit}, was varied between measurements, with \(A_{\text{read}} \in [min, max]\). The amplitude was initially set to and gradually increased up to . and the period of the cycles was \(T_{\text{read}} = 1\times10^{-3} s\). Each cycle was sampled uniformly in time, with 160 samples per period, totalling 16000 VI samples per waveform measurement. Measurements for the same state were collected consecutively in time, with a few seconds of delay between measurements to ensure that any excessive metastability associated with the state could be identified (through a change in the state).

\subsection{Data Preprocessing}
\label{sec:data_preprocessing}

Following the collection of measurements, the data was carefully reviewed and manually filtered to remove measurements where the state of the state of the device after SET had clearly changed significantly, apparent from observing a significant cycle-to-cycle difference in the Lissajous VI figures.
The remaining measurements included a wide range of state values and signal amplitudes and served as a dataset for evaluating our proposed model.

Following initial filtering, we used a straightforward minimum absolute-value iterative algorithm to attempt to detect the existence of non-zero starting point for the signals which can be caused by the trigger point of the probe being misaligned, removing \(N_{\text{discard}}\) points leading up to this point. We then discarded a number of values at the end of the series equal to \(N_{\text{period}} - N_{\text{discard}}\), in order to make the number of datapoints in each measurement the same, with \(N_{\text{period}}\) being the number of datapoints in each period.

Following the alignment of the periods, we use an automated fitting procedure to attempt to eliminate systematic measurement offsets for the voltage and the current values. This is done through minimising the magnitude of the quantity \(v \cdot i\) for points which lie in the upper left and bottom right quadrants of the VI plane.

The voltage across the memristive device is estimated as \(V_{t} - V_{s}\) and the current across the device as \(V_{s}/R_{s}\). Then the average of the voltage inputs of the series is calculated to determine if there is a systematic offset in the voltage magnitude. If so, then it is removed and the current is adjusted accordingly.

The small-signal memristor model (assuming the voltage and current do not have a significant impact on changing the state) can be calculated using the following equations:

\begin{align}
i_{mem} &= \frac{V_s}{R_s} \\
v_{mem} &= V_t - V_s \\
r_{mem} &= \frac{v_{mem}}{i_{mem}} = R_s \cdot \frac{V_t - V_s}{V_s} = R_s \cdot (\frac{V_t}{V_s} - 1)
\end{align}

, where $v_{mem}$, $i_{mem}$ and $r_{mem}$ are the memristor voltage, current and resistance respectively.

\subsection{Model Parameter Selection}

Following collection of the data and preprocessing, we attempted to fit a number of models to the data for comparison.

We adopt a grid search approach to identify the optimal model parameters for each model that we test. We conduct the grid search using a logarithmic scale to allow for searching over a wide range of parameter magnitudes, and gradual refinement over each iteration.

Given our set of model parameters represented by the vector \(\Theta = [G_m, \alpha_1, \alpha_2, \beta_1, \beta_2]\), we start with a set of upper and lower boundary values, given by the vectors \(B_{\text{lower}}\) and \(B_{\text{upper}}\), corresponding to the different parameters, and take a number \(n\) of equally logarithmically spaced values for each parameter in the range, including both boundary values as candidate values. For our experiments, we set \(n = 7\) and the boundary values for each parameter to be \(B_{\text{lower}} = 10^{-6}\) and \(B_{\text{upper]}} = 10^{2}\).

For each combination of parameter values, we approximate the best fit values of the state variable \(x\) for each VI set using the the Levenberg–Marquardt damped least-squares method~\cite{marquardtAlgorithmLeastSquaresEstimation1963} for every measurement by Equation~(\ref{eq:modelling_equation}). The model error is calculated according to a modified \ac{MSE} approach, described in Section~\ref{sec:fitting_error}.

For the optimisation step \(i\), following the determination of the set of parameters (point) with the least error, which we denote \(\Theta_i\), we adjust our boundary values by decreasing the size of the search for each parameter to be \(w = \log_{10}\left(\frac{B_\text{upper}}{B_{\text{lower}}}\right)\), setting the new boundary points for each parameter to be \(10^{(\log_{10}\left(\Theta_i\right) \pm w/2)}\) and selecting a new set of candidate points by interpolating logarithmically for \(n\) points once again.

We perform this iterative procedure \(m = 10\) times, selecting \(\Theta_m\) as our final optimised set of parameters for that particular model.

\subsubsection{Fitting Error}
\label{sec:fitting_error}

In choosing model parameters, we seek to minimise the fitting error of the model. One issue is that the fitting error will not take on the same magnitude for all output current magnitudes; said another way: if we imagine the fitting error to be a statistic of interest modelled as a conditional random variable, the random variable is not independent of the magnitude of the output.

As such, we must choose an appropriate metric for evaluation of the goodness of fit of our model to the acquired data. We choose to assume that the fitting error can be well approximated to be approximately independent of the magnitude of the current over a small current range. Thus, we choose to divide our current outputs into \(M\) sets of currents, each denoted \(\mathcal{I}_j\) for \(j \in 1 \ldots M\), over which we independently measure the fitting error, calculating the average fitting error as the mean weighted error, where we normalise the estimated error for each interval according to the given centroid of each current region, denoting the centroid for the set \(\mathcal{I}_j\) as \(C_j\). 

Given sets of data \(\mathcal{D}_i \in \mathbb{R}^2\), consisting of pairs of input voltages and output currents represented as 2-dimensional vectors, we can partition the set into subsets \(\mathcal{D}_{ij} = \{d \in D_{i} \; | \; d[0] \in \mathcal{I}_j\}\) - where \(i\) indexes the sets of data and \(j\) indexes the current regions.

This normalisation by the magnitude of the dependent variable is introduced to attempt to balance the differences in the average magnitude of the errors for different output magnitudes, based on the assumption that the fitting error will be approximately proportional to its magnitude (in turn, approximated by each centroid).
Division by the output current amounts to maximising the \ac{SNR} of the model prediction for each current region, averaged over the different regions, where the signal is the output current and the noise is the fitting error. We thus define the total fitting error - denoted by \(\mathcal{L}_i\) - for a test model \(f(\cdot)\) and a single data set \(\mathcal{D}_{i}\) as follows, with \(l\) being an arbitrary function that modifies the scaled error:

\begin{align}
    \label{eq:voltage_regions_error}
    \mathcal{L}_i = \frac{1}{M} \sum_{j} \frac{1}{|\mathcal{D}_{ij}|} \sum_{d \in \mathcal{D}_{ij}} l\left(\frac{\left|f(d[0]) - d[1]\right|}{\left|C_j\right|}\right)
\end{align}
and the overall fitting error, \(\mathcal{L}\), is calculated as the average of the fitting errors over all \(D\) sets of data:

\begin{align}
    \mathcal{L} = \frac{1}{D} \sum_i \mathcal{L}_i
\end{align}

For our experiments, we set \(l\) to be a quadratic function: \(l(x) = x^2\) such that the loss function is smooth.

In order to choose voltage regions, we use clustering methods based on the currents to group the output currents into similar-magnitude clusters. The KMeans algorithm with \(k = 8\) was used. In this specific context, the data to be clustered is one-dimensional and exhibits a monotonic increasing trend.

Division by the magnitude of the output current in the error calculation in Equation~\ref{eq:voltage_regions_error} is performed according to our assumption that the magnitude of the error in calculation of the output current from the input voltage is approximately proportional to the output current magnitude, hence division by the output current magnitude removes this proportionality.
Our approach is equivalent to calculating a statistical estimate of the error independently for each voltage region, by dividing by the number of samples in that region for each region. This is done in order to balance the error calculation, such that each voltage region contributes equally to the final error value, rather than letting the distribution of points across voltage regions affect which voltage regions have more of an effect on the final error.

\subsubsection{Evaluation Metrics}
\label{sec:eval_error}

In order to evaluate and compare the performance of our model to the \ac{Generalised MSS} and the modified \ac{Generalised MSS} models, we use the fitting error described above, in addition a number of other modifed errors, which can be characterised by changing the function \(l\).

In each case, we still scale the error magnitudes using the voltage region approach described in Section~\ref{sec:fitting_error}. The four evaluation metrics used are the \ac{MSE} (Equation~\ref{eq:mse}), \ac{MAE} (Equation~\ref{eq:mae}), and two other metrics based on the error ratio which we term the \ac{MRE} (Equation~\ref{eq:mre}) and the \ac{MRSE} (Equation~\ref{eq:mrse}). Note that we use a pedestal, \(\epsilon_1 = 1\times10^{-3}\) in the \ac{MRE} to avoid over-representing smaller values in the error, and a similar \(\epsilon_2 = 1 \times 10^{-6}\) in the \ac{MRSE}.

\begin{align}
    l_{\text{mse}}(y, \hat{y}) &= \mathbb{E}\left[\left(y - \hat{y}\right)^2\right]\label{eq:mse}
    \\
    l_{\text{mae}}(y, \hat{y}) &= \mathbb{E}\left[\left|y - \hat{y}\right|\right]\label{eq:mae}
    \\
    l_{\text{mre}}(y, \hat{y}) &= \mathbb{E}\left[\frac{\left|y - \hat{y}\right|}{|y| + \epsilon_1}\right]\label{eq:mre}
    \\
    l_{\text{mrse}}(y, \hat{y}) &= \mathbb{E}\left[\frac{\left(y - \hat{y}\right)^2}{(y)^2 + \epsilon_2}\right]\label{eq:mrse}
\end{align}

\section{Results}
\label{sec:results}

\subsection{Model Fitting Results}

Following our collection of the data and fitting of the model parameters according to the aforementioned procedure, we obtained the following parameters in Table~\ref{tab:res}:

\begin{table}[ht]
\centering
\caption{Estimated parameters for the discussed models.}
\label{tab:res}
\begin{tabular}{ccccc}
\multicolumn{5}{c}{GMSS}\\
$\alpha_1$ & $\alpha_2$ & $\beta_1$ & $\beta_2$ & $G_m$ \\
\hline
$2.730{\cdot}10^{\text{-}3}$ & $1.313{\cdot}10^{\text{-}7}$ & $1.392{\cdot}10^{1}$ & $2.327{\cdot}10^{\text{-}6}$ & $4.207{\cdot}10^{0}$  \\
\hline
\hline
\\
\multicolumn{5}{c}{Modified GMSS}\\
$\alpha_1$ & $\alpha_2$ & $\beta_1$ & $\beta_2$ & $G_m$ \\
\hline
$1.063{\cdot}10^{\text{-}2}$ & $3.869{\cdot}10^{\text{-}7}$ & $9.397{\cdot}10^{0}$ & $6.666{\cdot}10^{\text{-}9}$ & $3.160{\cdot}10^{0}$  \\
\hline
\hline
\\
\multicolumn{5}{c}{Proposed}\\
$\alpha_1$ & $\alpha_2$ & $\beta_1$ & $\beta_2$ & $G_m$ \\
\hline
$2.622{\cdot}10^{\text{-}1}$ & $6.597{\cdot}10^{\text{-}2}$ & $1.370{\cdot}10^{1}$ & $1.005{\cdot}10^{1}$ & $8.679{\cdot}10^{0}$  \\
\hline
\hline
\end{tabular}
\end{table}

\subsubsection{Comparison to other Models}

We report results for the proposed model alongside two other versions of the \ac{Generalised MSS}, chosen based on our discussion in Section~\ref{sec:state_modelling}.

The first of these is the original \ac{Generalised MSS}, given in Equation~\ref{eq:gmss_characteristic}, using diode Equation~\ref{eq:gmss_diode}, setting \(\alpha_1 = \alpha_2\) to ensure zero-crossing and avoid redundant complexity in the fitting procedure. We refer to this model as \ac{Generalised MSS}.

Secondly, we modify the \ac{Generalised MSS} diode equation to be the same as our proposed, more realistic, version given in Equation~\ref{eq:schockley_modified}, which frees the forward and reverse diode components to be independent. This is a model that therefore only includes the first modification proposed in Section~\ref{sec:state_modelling}. We refer to this model as modified \ac{Generalised MSS}.

Finally, we report results for the model with both the first and second modifications - the proposed model - which we refer to as the ``Proposed Model''. %

\subsubsection{Fitting Error}

Fitting error results for the different models, according to the metrics specified in Section~\ref{sec:fitting_error}, are shown in Table~\ref{tab:fitting_error}.

\begin{table}[ht]
\centering
\caption{\label{tab:res2} Fitting errors for the different models.}
\label{tab:fitting_error}
\begin{tabular}{c|cccc}
Model & \ac{MSE} [${\times}10^{\text{-}15}$] & \ac{MAE} [${\times}10^{\text{-}8}$] & \ac{MRE} & \ac{MRSE} \\
\hline
GMSS & $1.868$ & $3.152$ & $0.2325$ & $0.3240$ \\
\hline
Mod. GMSS & $2.040$ & $2.884$ & $0.1093$ & $0.06678$ \\
\hline
Proposed & $0.6507$ & $1.747$ & $0.08690$ & $0.05542$ \\
\hline
\end{tabular}
\end{table}

\subsubsection{Visual Model Fitting Results}

\begin{figure*}
    \centering
    \hfill
    \begin{subfigure}[t]{0.32\linewidth}
        \includegraphics[width=\linewidth]{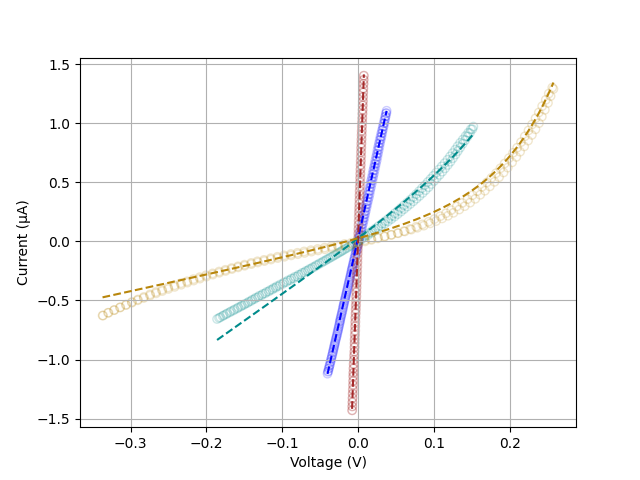}
        \caption{The GMSS model with fitted parameters.}
        \label{fig:GMSS_result}
    \end{subfigure}
    \hfill
    \begin{subfigure}[t]{0.32\linewidth}
        \includegraphics[width=\linewidth]{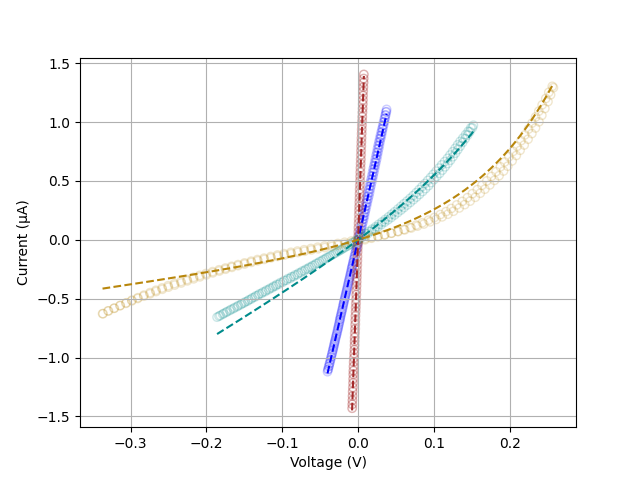}
        \caption{The modified GMSS model with fitted parameters.}
        \label{fig:GMSS2_result}
    \end{subfigure}
    \hfill
    \begin{subfigure}[t]{0.32\linewidth}
        \includegraphics[width=\linewidth]{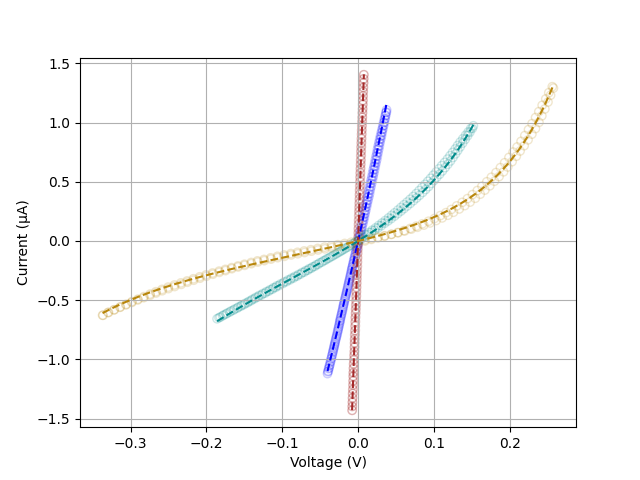}
        \caption{The model's fit for a subset of the measurements in the data collected, demonstrating a good visual fit to the asymmetric and exponential characteristics for a range of state magnitudes.}
        \label{fig:proposed_model_result}
    \end{subfigure}
    \caption{Fitting results for our proposed model (Figure~\ref{fig:proposed_model_result}, alongside that of the GMSS model (Figure~\ref{fig:GMSS_result}) and the modified GMSS model which includes only the first proposed modification to the diode equation (Figure~\ref{fig:GMSS2_result}). Our proposed model has a superior fit to that of the original GMSS and the modified GMSS, as visible in the figures and evident from the fitting errors in Table~\ref{tab:fitting_error}.}
    \label{fig:model_fitting_results}
\end{figure*}

The fit of the GMMS, modified GMMS, and ``Proposed Model'', to a range of experimental measurements from the dataset can be seen in Figure~\ref{fig:model_fitting_results}.

\subsection{Discussion}

The data collected as part of our experiments in Section~\ref{sec:data_preprocessing}, confirmed that the state of the \ac{SDC} memristive devices could not be accurately described by the instantaneous ratio of the voltage to the current. As such, our approach required a more complex physical model to describe the relationship between the voltage and the current, to characterise the state. It should be noted that our approach relies on the assumption that there the state of the memristor is Markovian.

The introduction of the state-dependence to the diode component of the current equation vastly improved the modelling results, compared to the state-independent approach originally proposed by Molter \& Nugent. This can be explained by the fact that changes to the memristive state will have an effect on the characteristics of the Schottky diodes at the interfaces as well, modulating the width of the barriers, for example, and our results suggest that this relationship is approximately linear. Further exploration of this effect and the connection between changes in the Ohmic and Schottky diode components of the IV characteristic could be useful to explore as part of future work.

\section{Empirical State Estimation}
\label{sec:empirical_state_estimation}

Given our model and its estimated parameters, may wish to estimate the state of the memristor, parameterised by \(x\), from future empirical measurements.

Our model of the relationship between the  voltage and current at a particular time, \((v, i)\), was given in Equation~\ref{eq:modelling_equation}. This can be rearranged to give a state estimate:

\begin{align}
    x = g(v, i) = \frac{i}{G_m v + \alpha_1 (e^{\beta_1 \cdot v} - 1) + \alpha_2 \cdot (1 - e^{-\beta_2 \cdot v})}
    \label{eq:state_estimate}
\end{align}

However, voltage and current measurements used to estimate the state in this way may be noisy. As such we propose a method producing an estimate which minimises the statistical uncertainty given noisy voltage and current measurements.

\subsection{Measurement Variance}

Based on observations of the noise present in the measurement equipment, for the purposes of quantification of uncertainty, we that the measurements acquired have a constant additive noise, which we model as a normal random variable.

We assume that our readings of the applied voltage and the voltage measured across the resistor are measured with a constant additive noise \(N\), with this noise being correlated between measurements taken simultaneously, accounting for an unpredictable systematic offset in the measurements:

\begin{align}
    \hat{v}_{\text{applied}} = v_{\text{applied}} + N \\
    \hat{v}_{\text{resistor}} = v_{\text{resistor}} + N
\end{align}

where

\begin{align}
    N \sim \mathcal{N}(0, \sigma_N^2)
\end{align}

We assume that the generator voltage \(v_{\text{applied}}\), which has a large magnitude and we apply to the series combination of the memristor and resistor, is relatively accurate, with any significant uncertainty arising from the measurement of voltage in \(v_{\text{measured}}\).

We obtain the voltage across the memristor as \(v_{\text{applied}} - v_{\text{resistor}}\) and we obtain the current across the memristor (and resistor) as \(v_{\text{resistor}}/R_{\text{series}}\).

The noisy measurement of the voltage across the memristor, \(\hat{v}_{\text{memristor}}\), is then approximately noiseless:

\begin{align}
    \hat{v}_{\text{memristor}} = v_{\text{memristor}} + N - N = v_{\text{memristor}}
\end{align}

And the noisy measured current \(\hat{i}_{\text{memristor}}\) is effectively:

\begin{align}
    \hat{i}_{\text{memristor}} = i_{\text{memristor}} + \frac{N}{R_{\text{series}}}
\end{align}

Our fit model, \(f(\cdot)\), which takes an implicit state \(x\) for the memristor, defining the relationship between \(v\) and \(i\) also gives us a corresponding model of the state variable \(x\) as a function of the values \(i\) and \(v\), which we can write as \(g(\cdot, \cdot)\). When \(g\) is applied to an IV pair, the resultant state estimate will thus have a noise with a variance that has a relationship to the magnitude of the voltages and currents applied. We can thus express the noisy state estimate of the model, \(\hat{\mathcal{E}}_f\), as follows:

\begin{align}
     \hat{\mathcal{E}}_f = \left(g\left(v_{\text{memristor}}, i_{\text{memristor}} + \frac{N}{R_{\text{series}}}\right)\right)
\end{align}

Assuming that the noise magnitude is small we can approximate \(g\) at point close to the IV pair using a first-degree Taylor series approximation at the point \(v_{\text{measured}}, i_{\text{measured}}\):

\begin{align}
    &g(v_{\text{memristor}}, i_{\text{memristor}} + \frac{N}{R_{\text{series}}}) \\ \nonumber &\approx 
    g(v_{\text{memristor}}, i_\text{memristor}) + Z \\
    &\text{where } Z \triangleq g_i(v_{\text{memristor}}, i_{\text{memristor}}) \cdot \frac{N}{R_{\text{series}}}
    \label{eq:noise_first_order_approximation}
\end{align}

where \(g_y\) denotes the partial derivative of the function \(g\) with respect to the variable \(y\).
Note that this approximation will rely on having a relatively accurate estimate of the partial gradient with respect to the current. For our \(g\), the partial gradient is a function only of the voltage, which we assume to be relatively noiseless, so this approximation holds.

According to Equation~\ref{eq:state_estimate}, we can approximate the \ac{SNR} of a particular measurement according to the first-order approximation given in Equation~\ref{eq:noise_first_order_approximation}:

\begin{align}
g_i(v, i) &= \frac{1}{(G_m v + \alpha_1 (e^{\beta_1 \cdot v} - 1) + \alpha_2 \cdot (1 - e^{-\beta_2 \cdot v}))}\\
Z &= g_i(v, i) \cdot \frac{N}{R_{\text{series}}} \\
\therefore \mathbb{E}[Z_k^2] &\propto g_i^2(v_k, i_k)
\end{align}

\subsection{Minimum Variance Estimate}

If we have a collection of measurements that can be used to estimate the state variable, we must trade-off using multiple measurements to minimise the noise variance, with giving higher weightings to measurements that are less uncertain (lower variance), with the uncertainty being a function of the input voltage, which is dependent on the function \(f(\cdot)\). This is because multiplying by a constant value results in the signal power being multiplied by the square of the constant, but also results in the noise power being multiplied by the constant. In the case of measurements with equal noise variance, there is no need to allocate differing weightings to each measurement, but if the variances of the measurements are assumed to be heterogeneous, then we must apply appropriate scaling factors to achieve a minimum variance in the estimate.

Note that because we cannot know the exact value of the state, we must use the noisy estimate, in order to estimate the magnitude of the variance of \(Z\).

The optimal weightings for the measurements can be calculated through minimisation of the power of the noise in the resultant convex sum estimate of the state variable formed from the noises of each of the \(K\) IV pairs, each contributing a noise term \(Z_k\). This will be measured relative to the increase in the power of the signal given by taking multiple measurements (i.e. the \ac{SNR}):

\begin{align}
    \max_{{m_k|\sum m_k = 1}} \frac{(\sum_k m_k)^2}{\sum_k m_k^2 \mathbb{E}[{Z_k}^2]}
\end{align}

This maximisation problem is equivalent to the minimisation of the denominator, since the numerator is always equal to 1, according to the maximisation constraint. Thus, the problem has an optimal solution which can be verified by using the method of Lagrange multipliers. This is achieved when the scaling factor is such that:

\begin{align}
    m_k = \frac{\frac{1}{\mathbb{E}[Z_k^2]}}{\sum_{l} \frac{1}{\mathbb{E}[Z_l^2]} }
    \label{eq:optimal_weightings}
\end{align}

We can see that the absolute magnitude of the variance of the noise \(N\), when sourced from the device, is not important, since the weightings are normalised, and it is the relative weightings that matter, as is clear from Equation~\ref{eq:optimal_weightings}. Thus, we do not need to worry about the absolute magnitude of \(N\) or \(V_{\text{off}}\). Note that the parameters of \(g\) are the same as those obtained for the fitting of \(f(v)\).

Note that in computation of the state estimate, we exclude measurements with voltages or currents of a very low magnitude (less than 30\% of the maximum value), due to the fact that we cannot assume that the voltage measurements are completely noiseless, and therefore the estimate of the partial derivative with respect to the current may be inaccurate at low magnitude points, since small fluctuations can cause large percentage deviations. %

\subsection{Demonstration}

\begin{figure}
    \centering
    \includegraphics[width=\linewidth]{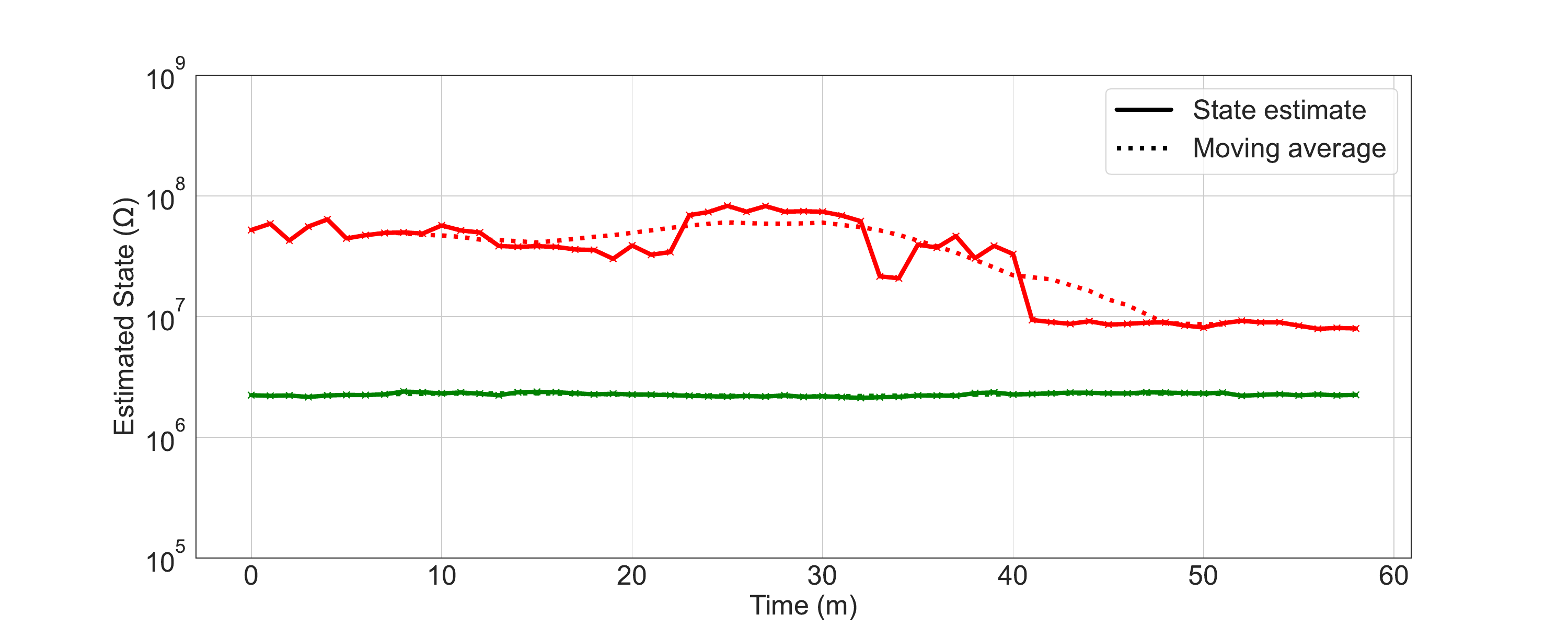}
    \caption{State estimation using the proposed model and the associated minimum variance state estimate, from various magnitude voltage and current measurements taken every minute for a period of 60 minutes. We demonstrate the state estimation procedure for two different initial states. The estimated state is given in Ohms by taking the inverse of the state variable for ease of interpretation, though the state is not an actual resistance due to the nonlinear VI relationship.}
    \label{fig:state_estimation}
\end{figure}

We demonstrate the state estimation procedure as applied to resistive drift data collected for the \ac{SDC} memristors. We characterise the devices by measuring their state every minute for 60 minutes following a set to two random initial states. Every minute, a READ waveform of duration 8 seconds is used to obtain VI samples that are to be used for state estimation. We demonstrate how the state estimation procedure for the proposed state estimation model exposes the dynamics of the state evolution over time, for the different initial states in Figure~\ref{fig:state_estimation}.

\section{Conclusions and future work}

In this work, we have investigated the use of one particular kind of memristive device for storage. We characterised the uniquely identifiable states in the device, according to mathematical modelling inspired by a proposed physical model of the device, demonstrating improvements over existing modelling approaches in accurately modelling the non-linear VI relation. We subsequently derived a minimum-variance approach for estimation of the proposed state variable from empirical data, in the presence of measurement error. The proposed state variable and associated measurement procedure can be used to model and capture the state of \ac{SDC} memristive devices for a range of emerging applications.

Future work may investigate the use of the proposed state characterisation protocol to evaluate the dynamics of \ac{SDC} memristors in a variety of contexts, for example, through measuring the energy cost associated with bringing the devices into different states as well as identifying resilient programming schemes which can most efficiently and reliably bring the device to a particular state. %

\begin{acronym}[]
    \acro{cGAN}{conditional GAN}
    \acro{CBRAM}{conductive bridge RAM}

    \acro{ECM}{electrochemical metallization}

    \acro{Generalised MSS}{Generalised Metastable Switch Model}

    \acro{JSCC}{joint source-channel coding}

    \acro{MEMS}{micro-electronic mechanical system}
    \acro{MRE}{mean relative error}
    \acro{MRSE}{mean relative squared error}
    \acro{MSE}{mean squared error}
    \acro{MAE}{mean absolute error}

    \acro{PCM}{phase change memory}

    \acro{SDC}{self-directed channel}
    \acro{SNR}{signal-to-noise-ratio}

    \acro{TCM}{thermochemical mechanism}

    \acro{VCM}{valence change mechanism}
\end{acronym}

\bibliographystyle{ieeetr}
\bibliography{references}

\end{document}